\documentclass[book,numbers]{elsbook}

\usepackage{amsmath,amssymb,amsfonts,amsthm,makeidx,graphicx}
\usepackage{txfonts}
\usepackage{helvet}
\usepackage{soul}

\newcommand{\fakeparagraph}[1]{\smallskip\noindent\textbf{#1.}}

\graphicspath{{Figs/}}

\makeindex

\begin{document}

\begin{frontmatter}

\chapter[Chapter Title]{Building Interoperable and Cross-Domain Semantic Web of Things Applications}



\author*[1]{Amelie Gyrard}%
\author[1]{Martin Serrano}%
\author[2]{Pankesh Patel}%

\address[1]{\orgname{Insight Centre for Data Analytics}, \orgdiv{IoT Unit}, \orgaddress{National University of Ireland, Galway, Ireland}}
\address[2]{\orgname{ABB Corporate Research}, \orgaddress{India}} 
\address*[1]{Corresponding: \email{amelie.gyrard@insight-centre.org}}

\titlemark{Managing the Web of Things: Linking the Real World to the Web}
\chaptermark{Chapter Title}

\makechaptertitle

\minitoc
\begin{abstract}[Abstract]              
The Web of Things (WoT) is rapidly growing in popularity getting the interest of not only technologist and scientific communities but industrial, system integrators and solution providers. The key aspect of the WoT to succeed is the relatively, easy-to-build ecosystems nature inherited from the web and the capacity for building end-to-end solutions. At the WoT connecting physical devices such as sensors, RFID tags or any devices that can send data through the Internet using the Web is almost automatic. The WoT shared data can be used to build smarter solutions that offer business services in the form of IoT applications. In this chapter, we review the main WoT challenges, with particular interest on highlighting those that rely on combining heterogeneous IoT data for the design of smarter services and applications and that benefit from data interoperability. Semantic web technologies help for overcoming with such challenges by addressing, among other ones the following objectives: 1) semantically annotating and unifying heterogeneous data, 2) enriching semantic WoT datasets with external knowledge graphs, and 3) providing an analysis of data by means of reasoning mechanisms to infer meaningful information. To overcome the challenge of building interoperable semantics-based IoT applications, the Machine-to-Machine Measurement (M3) semantic engine has been designed to semantically annotate WoT data, build the logic of smarter services and deduce meaningful knowledge by linking it to the external knowledge graphs available on the web. M3 assists application and business developers in designing interoperable Semantic Web of Things applications. Contributions in the context of European semantic-based WoT projects are discussed and a particular use case within FIESTA-IoT project is presented.

\end{abstract}

\begin{keywords}[Keywords:]
Internet of Things (IoT) \sep  Web of Things (WoT) \sep Semantic Web \sep Data Interoperability \sep Semantic Web of Things \sep Reasoning \sep Smart IoT \sep Programming Framework \sep Smart Services

\end{keywords}

\begin{points}[Chapter points]
\begin{itemize}
\item{} The evolution from IoT and Web of Things to Semantic Web of Things is explained.
\item{} Semantic web technologies applied to the Internet of Things domain are presented.
\item{} This chapter mainly focuses on extending the semantic part for the Web of Things.
\item{} We explain the proposed M3 framework and how to assist developers and system integrators in easily integrating semantic web technologies.
\item{} We focus on semantically annotating data, inferring meaningful information from WoT data and reusing knowledge expertise to build smarter WoT applications.

\end{itemize}%
\end{points}%
\end{frontmatter}%

\section{Introduction: Understanding Trends and the Evolution}
\label{chap1:sec1}

The \textbf{Internet of Things~(IoT)} vision is to connect sensors embedded into devices with the Internet and exploiting their services and functional capabilities \cite{aggarwal2013internet} \cite{vasseur2010interconnecting, patel-evalutioniot}. The multiple number of IoT applications are likely to revolutionize every aspects of our lives. For instance, Oral-B\footnote{http://connectedtoothbrush.com/, Last visited: September 2016} connected to the toothbrush controls dental hygiene. The Apple HealthKit\footnote{http://goo.gl/n2V42g, Last visited: September 2016} tracks fitness, nutrition and sleep.  However, the existing IoT applications largely focus on building dedicated scenarios and this limit the evolution of the IoT because each time new resources/devices are added, technical skills and ad-hoc adaptations are required using their own protocols and their proprietary data formats. Since such applications are already deployed and operating for a particular purpose, the main challenge is working on data interoperability in order to make the applications interoperable with each other. The main benefits are the possibility to compose simple applications to build more complex ones, but also to combine heterogeneous applicative domains to build innovative applications that co-exist with the already deployed ones.

On the other hand, in the last two decades the Web technologies have become very popular. In many areas, web solutions are the only option to make business, mainly because its potential capacity for ecosystem expansion of the number of users and because it is relatively simple to use, reliable and portable to multiple technology platforms (ie. smart phones, smart TVs, tablets, laptops, computers, etc). The most important aspect of the Web is the loose coupling between applications and computing servers. An example is HTTP that decouples a server and an application that accesses the server, the developer can change the functionality of the server without breaking the system. 
The \textbf{Web of Things (WoT)} has been already considered a part of the core activities in the Internet of Things -- \emph{leveraging what made the Web so successful and applying their principles to the physical devices} \cite{guinard2011internet, iswc2016tutorial}. The Web of Things is what makes possible that Internet of Things (IoT) developments and data can be accessible to a large number of Web developers and business designers and thus re-using the already available knowledge on the web to enhance IoT applications. System integrators and solution providers benefit from this accessibility and enable new and innovative cross-domain IoT applications. 


Recently, the combination of features and functional characteristics around the Internet of Things and the current demands in computing and processing capacity relay in the use of the cloud as the medium to solve interoperability issues, erroneously the cloud has been claimed as an interoperability facilitator, together with the Web technologies and the existing interoperability data services the cloud is making the IoT data manageable at the edge and particularly when IoT interoperability is required at the device level. This is called Fog of The Internet of Things or simply \textbf {Fog of Things} \cite{prazeres2016FogOfThings}. Fog of things aims for providing value to the data before making it available to the web facilitating the interoperability of the devices at the edge and preparing the managed data for further applications to be interoperable. This approach has demonstrated that the decoupling between physical devices and the web and bringing the computing capacity of the cloud down to the device is possible, explaining in details the Fog of Things is  out of the scope in this chapter but it is mention here as a matter of comprehensive summary and consideration in the IoT evolution.

Following the evolution, and the main objective of this chapter to identify methodologies for building cross-domain interoperability, not less important is the new trend following the integration of semantic web technologies to enhance the data and promote its use in multiple and diverse applications, also called \textbf{Semantic Web of Things (SWoT)} \cite{barnaghi2012semantics,swotsuiteframework} \cite{jara2014semantic}.  SWoT is independent of any domain, data can be generated by one domain and used in complementary domains. The best example here is an open data portal, where the data is offered to any developers to make use of it. The  data format is specified and mostly generic, thus multiple use applications can be developed. For example, a data set containing information about the number of available spaces in a car park, can be used used in different applications: for offering available spaces, but also for pre-booking or simply for estimating the occupancy of the place, etc. Semantic web technologies bring several benefits: 1) Semantically annotating sensor datasets to unify heterogeneous data and explicitly describe metadata, 2) enriching semantic sensor datasets with external knowledge graphs available on the Web to add value to them and most important operate them in a more knowledge-based manner, and 3) performing analytics on data by means of applied logic and reasoning mechanisms to deduce meaningful additional information from data. 

In this chapter, we will focus on this last trend, assuming that Internet of Things and Web of Things have resolved the challenging aspects of transparently sharing information with a defined format amongst devices. In the next section, we describe a simple SWoT application, more complex ones could have been included but the main intention is to simplify the understanding process and highlight advantages when using semantic web technologies. The scenario is relevant when we observe the added capabilities for cross-domain usage and interoperablity of the information. It is also the main objective to use this simple approach to illustrate characteristics and motivation for building interoperable and cross-domain SWoT applications based on those identified characteristics. Annotated and metadata aggregations are considered here as part of the methodologies to enable interoperability.

A thermometer could be plugged into the Web to retrieve data and an application on the Web will show statistical and analytical information to make the user of the thermometer aware of the conditions. Most of the existing Internet of Things applications would just enable the visualization of the data produced by the thermometer in the form of a ``dashboard`` including location on where and when data was collected, trends and perhaps some forecasting on what could happen based on historical and other external sensor data information (i.e. humidity, pressure, etc.) Beyond that non-special solution, the example in this case is the need for building an application that assists the humans to automatically interpret the visualized data and combines analytical and statistical data while reusing knowledge databases designed by specialist experts. This mean the usage of healthcare databases to define if the temperature reported by the thermometer located in a room is affecting the body temperature of the people. For instance, people having asthma will be more prone to have problems if the temperature is not controlled adequately. The Web of Things will work to have all the thermometers ``talking`` with each other and via web applications. It will also simplify the decision making of the building manager by taking the most optimal average temperature according to pre-defined conditions such as period of the year (e.g., summer, winter, spring) or the time of the day (e.g., morning, evening, etc.).
The Semantic Web of Things application would additionally help to identify the patterns that has been reported historically as problematics and based on specific periods and use of the data to correlate with current knowledge databases that report similar symptoms.
The developers using SWoT approach can design such applications to combine the data produced by the body thermometer with the healthcare knowledge databases. More services can be offered for example when there is detected an increase in the normal body temperature. Not only healthcare knowledge databases can be used but also additional ones to suggest causes and home remedies. This example shows the necessity of making not only two different domains interoperable: Healthcare and Food but it also highlights the need for reusing domain-knowledge expertise (e.g., models to structure data and reasoning mechanisms to add value to data) available on the Web.
This example emphasizes on integrating a reasoning mechanism specific to cross-domain knowledge databases but also the need to add value to data which is the main objective of using SWoT.

At this point, we have described the evolution and the main differences in the Internet of Things mayor trends, i.e. Internet of Things, Web of Things, Fog of Things and Semantic Web of Things. In the next section, we focus on describing application development requirements to build cross-domain Semantic Web of Things applications.


The remainder of  this chapter is organized as follows: Section~\ref{sec:related work and challenges identification} reviews the related work that applies Semantic Web to IoT and work towards identifying the most common challenges for enabling cross-domain interoperability. Moreover, a study about the limitations of the existing work and the justification about the need of a comprehensive framework for SWoT is explained. Section~\ref{sec:m3frmwrk} presents our contributions and the M3 framework that focuses on the semantic part of the WoT and which primary objective is for assisting developers in designing SWoT applications. Section~\ref{sec:summary} summarizes this chapter and describes briefly some future directions for the SWoT area and the refinement of the M3 framework.

\section{Related Work and Challenges Identification}
\label{sec:related work and challenges identification}
This section introduces the most relevant work that could be applied to WoT regarding data interoperability, its modeling and reasoning mechanisms integrating semantic web technologies and addressing the research challenges towards enabling SWoT interoperability. 


\subsection{Technical Requirements}
\label{sec:challenges} 

In this section, we review application development requirements as learned from the analysis of application examples, research studies~\cite{patel-jss, iotsuite-icdcn} and based on practical implementation of Internet of Things solutions and Web of Things modeling in the context of OpenIoT framework\footnote{http://www.openiot.eu/, Open Source Middleware for the Internet of Things} and VITAL platform\footnote{http://vital-iot.eu/, The future of connecting IoT smart city systems} and FIESTA-IoT portal\footnote{http://fiesta-iot.eu/, Federated Semantic interoperability for Internet of Things systems}. The projects have been coordinated and the implementations and developments lead by consortium technical team. We focus in providing a selection of challenges based on identified solutions on how to build cross-domain interoperability.


\fakeparagraph{Ensuring interoperability among heterogeneous data} Devices are not interoperable with each other since data is exchanged following non standardized protocols using proprietary data formats and they do not use common taxonomies or vocabularies~\cite{ipsn2017demo, iotsuite-icse2016}. Usually, IoT devices provide unformatted data names as ``raw`` sensor data. This ``raw`` sensor data does not contain any additional description or metadata and requires specialized knowledge and manual effort in order to build cross-domain applications.

\fakeparagraph{Deducing meaningful information from raw data} Users are primarily interested in real-world entities (such as people, places and things) and their high-level knowledge~(e.g., deriving snowfall from temperature and precipitation measurements, a body temperature is abnormal or not) rather than raw output data produced by sensors attached with these entities. 

\fakeparagraph{Reusing and integrating the domain knowledge already available on the Web to enable WoT data} Knowledge already available in the web can be used to simplify complex knowledge operations, for instance, knowledge bases already designed for smart homes could be reused to describe a smart home which comprises thermometers, smoke detectors, humidity sensors and their related actions to take. This reusability is crucial in the Web of Things in order to enable data interoperability.


\fakeparagraph{Ensuring interoperability among WoT projects} Taking inspiration from ``\textit{sharing and reuse}`` approaches, an effort should be done to reuse data, vocabularies, designs, and softwares already done in the past to encourage reusability but also to build composite and interoperable applications that uses annotated data for enabling the creation of more WoT services. There is a need to study the existing projects and find a common pattern of the components constantly redesigned (e.g., the model to structure data or the reasoning mechanisms).

\fakeparagraph{Combining different application domains}
Combining different domains could enable smarter applications. It requires an interoperable domain knowledge to easily navigate from one domain-specific knowledge graph to another. This approach takes inspiration from the Semantic Web community designing Ontology Design Patterns~(ODPs) \cite{gangemi2009ontology} and Ontology Networks \cite{suarez2010neon}.    \newline

In addition to the listed requirements, achieving such challenges would help to enable a more directional approach towards building interoperable solutions, best SWoT practices are founded to encourage replicability. The most important is to work on:

\begin{itemize}
\item \textit{Reducing the time spent for developing WoT applications}. In order to create interoperable and cross-domain SWoT applications, developers have to perform various tasks such as designing an application, semantically annotating data and interpreting data. To perform these tasks, developers have to learn semantic web technologies and tools, a time consuming process, which can take several months. Reducing this gap as much as possible can be done by empowering a framework that assist developers in designing interoperable applications without learning semantic web technologies~\cite{gpce2015, apsec2015}. 
\item \textit{Reducing the learning curve required by WoT developers to integrate semantic web technologies}.
Fast prototyping of semantic-based WoT applications by hiding the use of semantic web technologies as much as possible is required to avoid the developers burden on designing ontologies, semantic annotators and reasoning mechanisms to enrich their data. An extensive work with Web frameworks (e.g., Drupal, Wordpress) has been done to design pre-defined templates to automatically generate web sites to avoid users dealing with Web technologies. Based on this idea, pre-defined templates to design SWoT applications can be created. 
\end{itemize}

\subsection{Semantically Annotating Data}
\label{sec:annotation}
This section presents approaches that leverage the semantic web technologies for annotating data and achieve data 
interoperability.

\textbf{Semantic Sensor Web} is designed to semantically annotate sensor data with Semantic Web languages such as Resource Description Framework in Attributes (RDFa)~\cite{sheth2008semantic}. Semantic Sensor Web uses and defines ontologies to support interoperability over heterogeneous environments and to describe concepts and units related to applicative domain. It also introduced the idea to reason over semantic sensor data to infer new knowledge in two domain-specific scenarios: weather and healthcare. For instance, in the weather domain ``Potentially Icy``, ``Low Visibility`` and ``High Winds`` can be deduced. Semantic Sensor Web leads to a set of tools such as SemSOS~\cite{henson2009semsos} and IntelligO \cite{henson2013semantics}. SemSOS has been designed for accessing and querying sensor data on the web. SemSOS uses the 52° North's SOS\footnote{http://52north.org/communities/sensorweb/sos/} implementation and enriches the SOS service with semantic annotations. Both tools uses semantic web technologies to manage sensors measurements. 

\textbf{Linked Sensor Data} is an approach to semantically annotate the MesoWest weather dataset to publish a unified dataset on the web available as Linked Data~\cite{patni2010linked}. Linked Open Data (LOD) is an open-based sharing and reusing approach for publishing, sharing, reusing and combining data on the Web \cite{bizer2009linked}. It is based on Sensor Web Enablement~(SWE) \cite{botts2008ogc} standards to retrieve sensor measurements, convert data encoded with O\&M\footnote{O\&M stands for Observations and Measurements,  http://www.opengeospatial.org/standards/om} into RDF\footnote{RDF stands for Resource Description Framework, http://www.w3.org/TR/REC-rdf-syntax/}, and then publish semantic sensor datasets on the Web. The datasets comprise 20,000 sensors, 160 million sensor observations and 1.7 billion RDF statements. Datasets have been enriched with contextual information using the GeoNames dataset to deduce regions, etc. Real sensor datasets have been semantically annotated to design specific applications without having in mind application interoperability.


\textbf{Linked Sensor Middleware} is an open source middleware where more than 110,000 sensors from open data endpoints are included. LSM follows the
W3C\footnote{W3C stands for World Wide Web Consortium} SSN extended Group recommendations. The SSN ontology, \cite{compton2012ssn} is one of the main outcome of the extended group and was originally designed for agriculture, ocean observations, smart vineyard and smart farm scenarios which have been extensively used in IoT domains.  

\textbf{LD4Sensors} is an approach to semantically annotate sensor data coming from different platforms (e.g., from different weather stations) and link data together \cite{leggieri2011incontext}. This mechanism is called ``Linked Data for Sensors``  and enables aggregating same kind of data generated by heterogeneous devices using heterogeneous terms. The LD4Sensors approach also attaches explicit metadata to raw data by using semantic web technologies and enables publishing ``linking sensor data`` on the web, so other users can reuse it to build innovative applications.  

\subsection{Knowledge Discovery: Reusing Domain Knowledge}
\label{sec:knowledge}
To ease interoperability among WoT applications and services, reusing existing ontologies is highly encouraged. In this section, semantic search engines and ontology \& dataset repositories for reusing the background knowledge available on the Web are briefly studied and presented.

\subsubsection{Semantic Search Engines}
\label{sec:searchengine}
The \textbf{Neon} project provides ontology methodologies, and suggests semantic web search engines to find and reuse domain ontologies \cite{suarez2010neon}. Since the integration of semantic web technologies within IoT is emerging, frequently, a new ontology is redesigned instead of reusing the existing ones as preconized by Noy et al.: ``an ontology is designed to be shared and reused`` \cite{noy2001ontology}. Semantic search engines such as \textbf{Sindice} \cite{tummarello2007sindice}, \textbf{Watson} and \textbf{Swoogle} \cite{ding2004swoogle} cannot reference relevant domain ontologies for WoT since they are not published online most of the time. There is a real need to spread semantic web best practices within the IoT community. A comprehensive survey of the best practices for publishing, sharing and reusing IoT ontologies is summarized \cite{gyrard2015wfiotBestPractices}. For instance, tools to automatically document ontologies are referenced, and recommendations for adding ontology metadata and publishing the ontology online with a good namespace are given.

\subsubsection{Ontology and Dataset Repositories}
\label{sec:ontology}
\textbf{Datalift} \cite{scharffe2012enabling} is a project assisting people in semantically annotating and linking data, but not in IoT area and it does not provide any vocabularies relevant for IoT. Datalift provides the \textbf{Linked Open Vocabularies (LOV)} \cite{vandenbusschelov2015}, an ontology catalogue, mainly known by semantic web experts. LOV lacks of ontologies relevant for IoT, and does not accept new ontologies if they do not follow semantic web best practices. \textbf{DataHub} is a dataset catalogue and does not provide quality checking when submitting a new dataset which leads to interoperability issues when consumers want to reuse and combine datasets. 

\subsection{Deducing Meaningful Information from Data}
\label{sec:deduce}
To interpret IoT data, the existing approaches explained in this section are mainly based on machine learning based approaches such as clustering to extract useful information from data. However, as previously mentioned in Section \ref{sec:annotation}, some projects used semantic web technologies to annotate data. From a more practical point of view, it would be easier to use Semantic Web Rule Language (SWRL)\footnote{https://www.w3.org/Submission/SWRL/} and reasoning/inference engines to reason on data. 

\textbf{IntellegO} is a semantic perception approach to interpret and reason on sensor data \cite{henson2013semantics}. IntellegO uses an abductive logic framework and Parsimonious Covering Theory (PCT) to interpret data based on an ``ontology of perception``. The development and reuse of the background knowledge (i.e., domain knowledge designed by domain experts for instance in healthcare) required for interpreting data is a difficult task and is not considered in this work. 

\textbf{Knowledge Acquisition Toolkit (KAT)} enables pre-processing and cleansing of data to reduce the traffic in network communications \cite{ganzKat2014}. KAT is composed of three components: 1) An extension of Symbolic Aggregate Approximation (SAX) algorithm, called SensorSAX, 2) abductive reasoning based on the Parsimonious Covering Theory (PCT), and 3) temporal and spatial reasoning. KAT is based on machine learning techniques (k-means clustering and Markov model methods). However, KAT neither deduces meaningful information from sensor data nor exploits domain-specific background knowledge relevant for IoT, which is already available on the web. 

\subsection{Relevant Semantic-based WoT Projects}
\label{sec:projects}
In this section existing approaches and projects working on interoperability by integrating semantic web technologies to WoT platforms are described: Spitfire, OpenIoT, CityPulse, VITAL, and FIESTA-IoT.

\textbf{SPITFIRE} introduced the Semantic Web of Things concept to integrate semantic web technologies to the Web of Things \cite{pfisterer2011spitfire}. SPITFIRE connected the real world of things and the Web by proposing concepts, methods and software infrastructure. Spitfire provided tools such as smart-service-proxy, LD4Sensors for linking sensor data presented above, Web-based Task Assignment and execution (WebTAsX), visualizer server and gateway connection mapper.

\textbf{OpenIoT\footnote{http://www.openiot.eu/, Last visited: September 2016}} is an open-source IoT platform enabling the semantic interoperability of IoT services in the cloud \cite{soldatos2015openiot}. OpenIoT introduced the ``\textit{Sensing-as-a-Service}`` principle by converging cloud infrastructures with IoT applications facilitating IoT service creation and deployment. OpenIoT has been successfully demonstrated and implemented by means of using semantic web design and this is a major advancement over the IoT/cloud infrastructure state-of-the-art. Previously, the state-of-the-art was characterized by an essential lack of semantic interoperability and integration across the diverse IoT applications and sensor data streams. OpenIoT brings IoT applications, Semantic Web design and Cloud infrastructure technologies all together.
The OpenIoT platform is a joint effort of awarded open source contributions associated with popular RFID and Wireless Sensor Networks (WSN) such as the Global Sensor Networks (GSN) \cite{calbimonte2014xgsn}, and the Linked Sensor Middleware (LSM) \cite{lePhuoc2014GraphOfThings} to connect heterogeneous sensors for real-time processing. OpenIoT has been applied, but not limited, to four scenarios: smart cities by enabling crowd sensing for air quality analysis, smart agriculture for monitoring large-scale deployments, intelligent manufacturing for optimizing logistic processes and traceability of material, and the smart building in a university campus for indoor sensing and booking resources for specific activities associated with students study rooms. 

\textbf{CityPulse\footnote{http://www.ict-citypulse.eu/page/, Last visited: September 2016}} is mainly focused on large-scale analysis and real-time intelligence to extract meaningful knowledge and perceptions from heterogeneous data streams \cite{barnaghicitypulse}. CityPulse works on designing and developing a framework, which supports the development of 101 applications for smart cities (e.g., public parking space availability prediction). CityPulse contains tools which can be used for discovering, processing and interpreting the data sources (e.g. weather data, traffic data, etc.) and social data streams (e.g. Facebook, Twitter, Google) by using Complex Event Processing (CEP) techniques. CityPulse works on bridging the gap between the application technologies on the WoT and the real world data streams. 

\textbf{VITAL\footnote{http://vital-iot.eu/, Last visited: September 2016}} is a platform that focuses on reducing the development and deployment costs of smart city applications \cite{petrolo2015towards} \cite{petrolo2014integrating} by exploiting the functionality for interconnect legacy data with IoT sensor collected data by using semantics. VITAL has deployed demonstrators in two cities: Istanbul and London. VITAL platform reuses the work achieved in OpenIoT and the W3C SSN-XG in terms of ontologies. VITAL follows a specific model enabling for the first-time an operating system that works across the cities and provides software tools based on semantic annotation and complex event processing to integrate heterogeneous IoT systems (legacy) and ensures data integration, and interoperable IoT services. 

\textbf{FIESTA-IoT\footnote{http://fiesta-iot.eu/, Last visited: September 2016}}, stands for Federated Interoperable Semantic IoT/cloud Testbeds and Applications, is a project, that reuses the previous work done in European project such as OpenIoT, CityPulse, VITAL, and SmartSantander. The FIESTA-IoT project works on integrating IoT platforms, testbeds, data, and associated silo applications. FIESTA-IoT aims for opening up new opportunities in the development and deployment of experiments that exploit data and capabilities from multiple testbeds. The FIESTA-IoT infrastructure looks at enabling experimenters to use a single Experiment-as-a-Service (EaaS) API for executing experiments such as reasoning on sensor data or sensor data discovery. Such experiments are conducted over multiple IoT federated testbeds in a testbed agnostic way i.e. like accessing a single large scale virtualized testbed. 
The main goal of the FIESTA-IoT project is to open new horizons in the development and deployment of IoT applications and experiments at a global scale, based on the interconnection and interoperability of diverse IoT platforms and testbeds. FIESTA-IoT project's experimental infrastructure is targeting to be the entry point for European experimenters in the IoT domain with the unique capability for accessing to and sharing IoT datasets in a testbed-agnostic way. It enables the execution of experiments across multiple IoT testbeds, based on a single API for submitting the experiments.

\subsection{Limitations of Existing Approaches}
\label{sec:limitations}
Most of the existing WoT projects and tools described above are using Semantic Web principles and available technologies for integrating semantic-enabled WoT services. The state-of-the-art analysis reveals that interoperability issues remain open since existing projects constantly redesign their own models to structure data, semantic annotators, semantic based IoT data analytics mechanisms, and domain-specific IoT applications. Based on this study, important notes can be addressed such as follows. There are new methods inspired from the Linked Open Data (LOD) approach to reuse existing efforts in order to help WoT developers to interpret IoT data and ease interoperability among applications. To ensure interoperability, the reuse of efforts and background knowledge already designed is highly encouraged.
The following research directions regarding reusing background knowledge are highlighted: 1) Extracting domain knowledge (ontologies, datasets and rules), 2) combining domain knowledge, 3) new ontology mapping, alignment and merging tools adapted to IoT ontologies, 4) make the domain knowledge interoperable by using semantic web methodologies and best practices, 5) integrate a semantic reasoning engine and reuse ``IF THEN ELSE`` rules already designed, and 6) an assistance for IoT application development using semantic web technologies. Overcoming such challenges will enable providing interoperability among semantic-based WoT applications and providing a uniform guideline to the application developers.

\section{Contributions and M3 framework}
\label{sec:m3frmwrk}

We propose a comprehensive Machine-to-Machine Measurement~(M3) framework that provides the entire data workflow generated by devices to build SWoT applications. This framework creates meta data models and semantically annotates sensor data, deduces meaningful information out of data, reuses the domain knowledge available on the Web. More specifically, the M3 addresses the semantic-related requirements, discussed in Section~\ref{sec:challenges}, as follows: 


\begin{itemize}
\item ``Ensuring interoperability among heterogeneous data`` is addressed by using a taxonomy to unify terms and deal with issues such as synonyms and abbreviations, thus semantic web technologies for \textit{unifying} data are used.

\item ``Deducing meaningful information from raw data`` is addressed  by using reasoning mechanisms such as logic-based inference engine (e.g., if temperature is 38$^\circ$C then hot).

\item ``Reusing and integrating the domain knowledge already available on the Web to enrich WoT data`` is addressed by classifying domain knowledge as a new dataset.

\item ``Ensuring interoperability among WoT projects`` is addressed by ensuring interoperability among data, models to structure data and reasoning mechanisms.

\item ``Combining different application domains`` is addressed by generating rules and use them for domain knowledge control.

\end{itemize}

\begin{figure}[!htbp]
\centering
\FIG{\includegraphics[width=\linewidth]{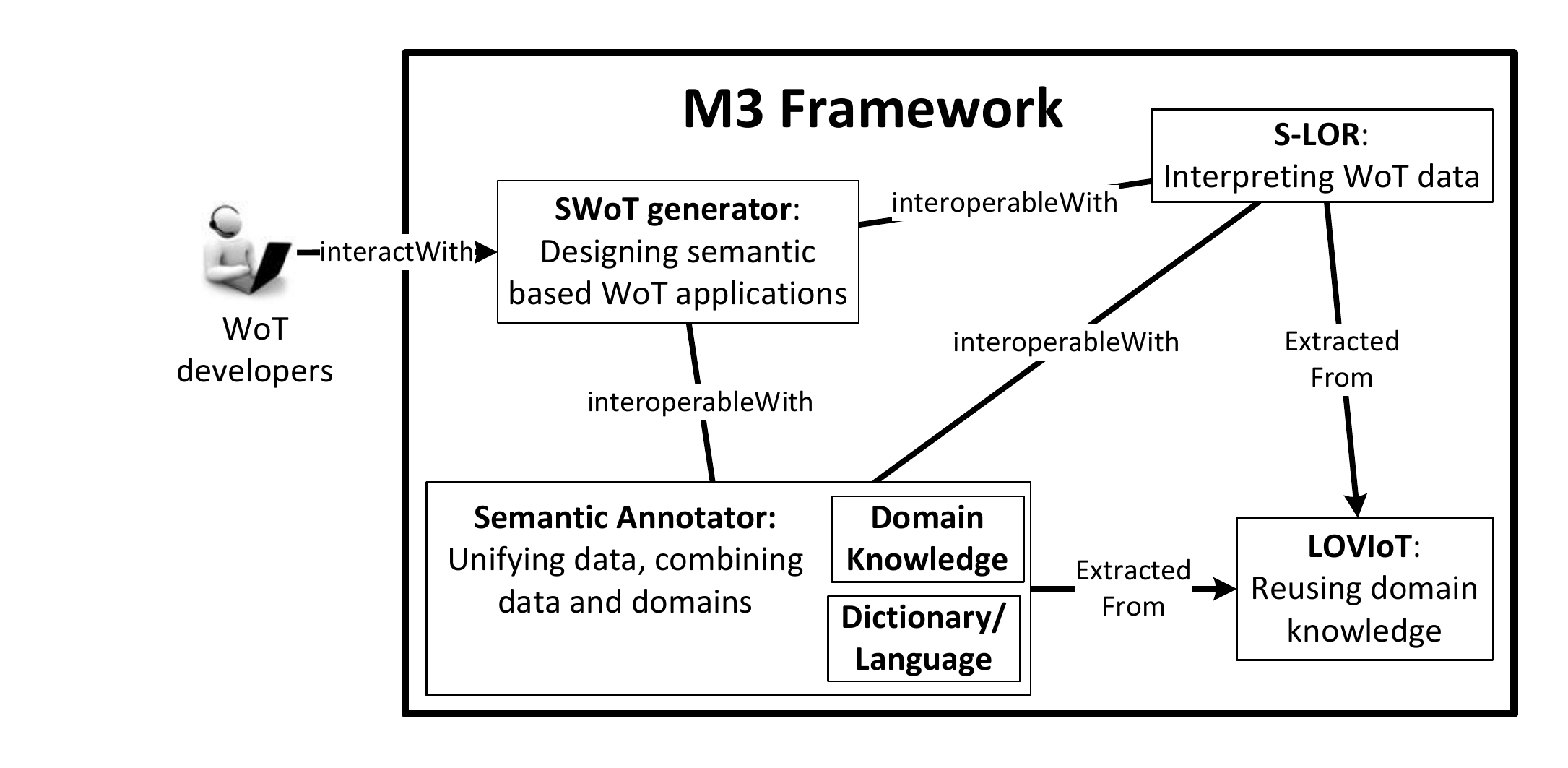}}
\caption{The M3 Framework and functional relations}
\label{M3FrameworkChallenge} 
\end{figure}

To overcome the limitations of the WoT enabling semantic interoperability, the Machine-to-Machine Measurement (M3) framework has been designed to assist developers in designing semantic WoT applications. ``Measurement`` explains that M3 is mainly focused on data interoperability. M3 is a framework, a semantic engine and a methodology to define an entire workflow exploiting data produced by devices to: 1) Semantically annotate data, 2) reason over data to infer new knowledge, and 3) provide Semantic Web of Things (SWoT) templates to ease the task of WoT developers who are not familiar with semantic web technologies to easily develop semantic-based applications. The M3 framework and its four sub-components explained thereafter addresses the following challenges (introduced in Section \ref{sec:challenges}), as depicted in Figure \ref{M3FrameworkChallenge}:  



\subsection{A semantic Engine for WoT}
\label{sec:dataworkflow}

\begin{figure}[!htbp]
\centering
\FIG{\includegraphics[width=\linewidth]{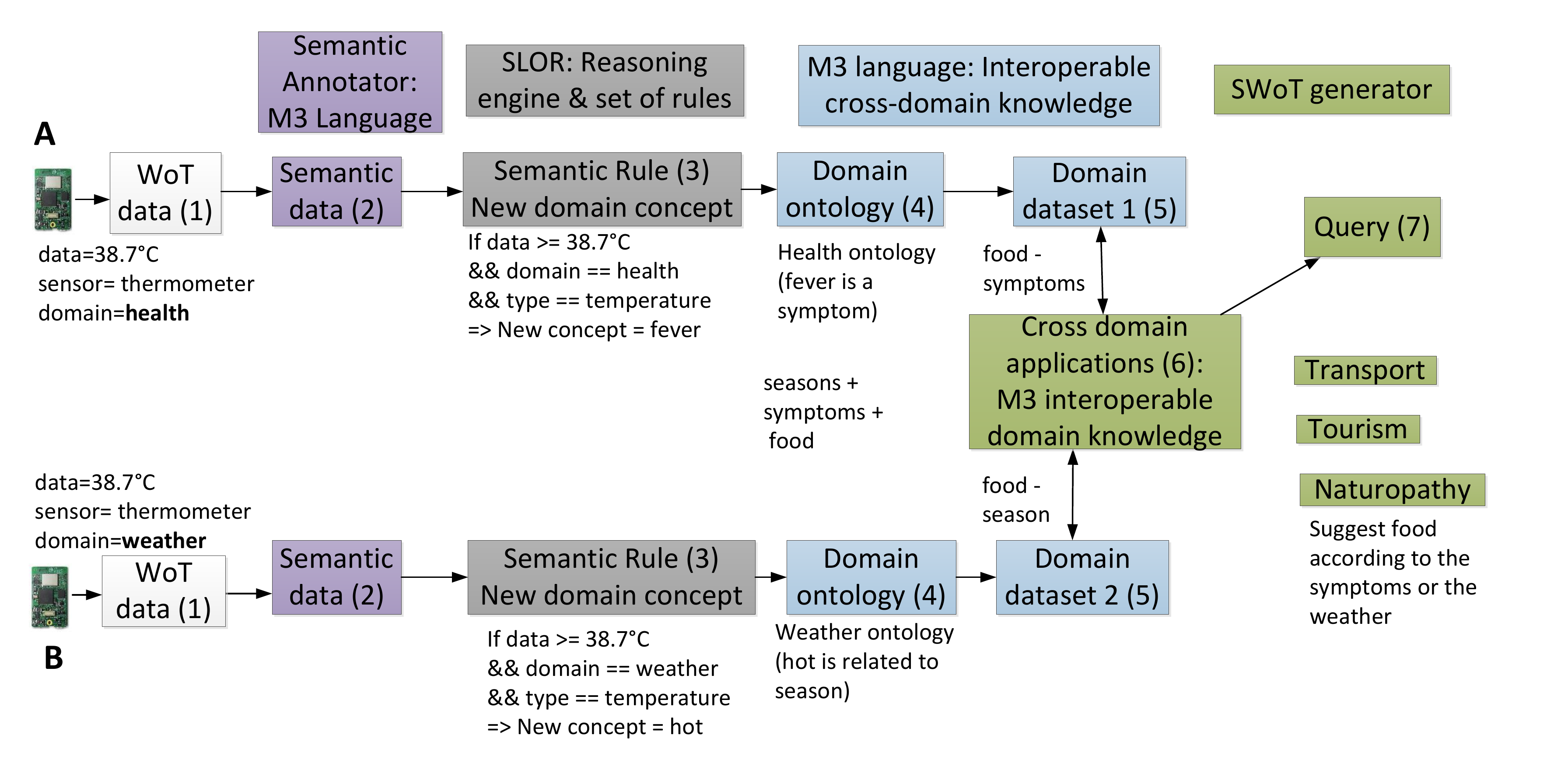}}
\caption{The M3 Semantic Engine Workflow}
\label{M3SemanticEngine} 
\end{figure}

The figure \ref{M3SemanticEngine} summarizes the M3 semantic engine workflow \cite{gyrard2014wfiot}, and at the same time it maps the use case described in the introduction of this chapter showing same data (e.g., temperature 38.7$^\circ$C) produced by a thermometer from two different applicative domains: Path A (upper side in the figure) for healthcare and path B (lower side) for weather forecasting. This example highlights the necessity to 1) Explicitly add description to sensor measurements, 2) interpret data, and 3) combine domains to design cross-domain applications. The first box, called ``WoT data`` returns sensor descriptions such as temperature 38.7$^\circ$C. In the second box, called ``Semantic data``, previous data is semantically annotated according to the M3 language, implemented as an ontology, which is required for the future steps. 

In the box called ``Semantic Rule, new domain concept``, the S-LOR approach is exploited, and a set of interoperable rules compliant with the M3 language to infer new knowledge used. In path A, S-LOR deduces the concept ``fever``, whereas in path B, S-LOR deduces the concept ``hot``. In the boxes called ``Domain ontologies`` and ``Domain datasets`` the results of the reasoning provided by S-LOR are linked to the M3 interoperable domain ontologies and datasets used in the SWoT templates. Such interoperable domain knowledge has been extracted from the LOV4IoT dataset. In step 6), ``Cross domain applications``, the M3 interoperable domain knowledge is used to combine domains and provide suggestions. For instance, food related to the fever symptom in path A, and food related to season in path B. Since food referred to the same namespace in both domain knowledge, it is easy to combine domains. Finally, in step 7), a request queries the M3 interoperable cross-domain knowledge to get smarter data and suggestions as result of a SPARQL query for example. All of these steps can be done by loading the SWoT template provided by the SWoT generator and a Java skeleton that we provide\footnote{\url{http://sensormeasurement.appspot.com/?p=end_to_end_scenario}} to easily build semantic-based IoT applications and enrich IoT data. The provided results will be later parsed and exploited in the final application such as the naturopathy application which suggests home remedies when a high temperature (which could be a potential fever) is detected. The final application could be a user-friendly interface or could even send notifications, alerts or could send order to actuators.

\subsection{Main Functionality and Semantic Value} 


In the following section, we describe essential functionalities for enabling data interoperability in the context of using the M3 framework and its semantic orientation towards data interoperability. The cross domain nature comes across as result of following the requirements described in previously introduced in Section~\ref{sec:challenges}. 

\begin{itemize}
	\item \textbf{SWoT generator}, produces templates to the developers to ease their task in development. The template contains the files required to execute the semantic functionality and passes this to the semantic annotator and SLOR. SWoT is explained in detail in section \ref{sec:swotgenerator}, 
    
    \item \textbf{Semantic Annotator}, explained in section \ref{sec:m3language}, comprises a dictionary/language to semantically annotate data in a unified way compliant with the M3 framework, a cornerstone component for the well execution of the S-LOR reasoning engine.
    
    \item \textbf{Sensor-based Linked Open Rules~(S-LOR)}, explained in Section \ref{sec:slor}, is the inference engine executing the dataset of interoperable rules to deduce meaningful information from semantic data.
    
    \item \textbf{Linked Open Vocabularies for Internet of Things~(LOV4IoT)}, explained in Section \ref{sec:lov4iot}, is a dataset and a set of APIs to retrieve domain knowledge required. This domain knowledge has been manually redesigned to be interoperable and used by the M3 language, SLOR and SWoT generator.
    
\end{itemize}

 

In this section, we present the M3 framework, and explained its functional relations .

\subsection{SWoT Generation}
\label{sec:swotgenerator}
The SWoT generation or SWoT Generator is a necessary process in any SWoT Application with the objective to ease the task of WoT developers in designing semantic-based WoT applications \cite{gyrard2015swotgenerator}. The SWoT generation provides the domain knowledge required to build semantic-based WoT applications by reducing development costs. 
Developers can interact with SWoT generator though Application Programming Interface (API) or Graphical User Interface (GUI).

\begin{figure}[!htbp]
\centering
\FIG{\includegraphics[width=\linewidth]{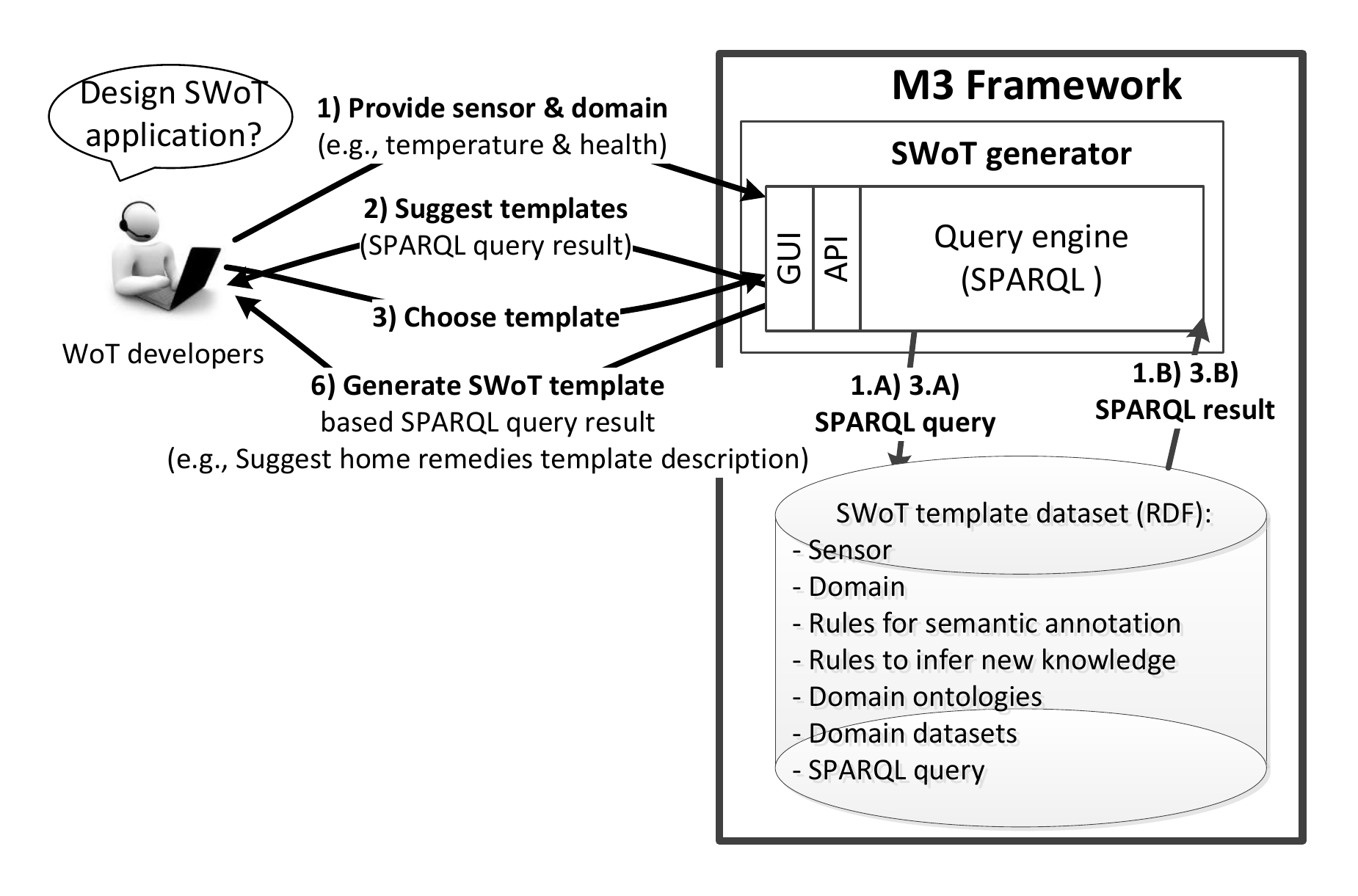}}
\caption{The SWoT generator and its semantic-based IoT templates generation cycle}
\label{SWOTgenerator} 
\end{figure}

Figure \ref{SWOTgenerator}, shows that developers interact with the SWoT generator by providing sensors used within the IoT application that they want to develop and in which domain the sensors are deployed (e.g., thermometer in healthcare). The SWoT generator interacts with a RDF template dataset through SPARQL queries and returns several SWoT templates matching the requirements. The template dataset has been manually designed and implemented with RDF. The current dataset contains 33 templates to design semantic-based applications. The developers choose one template (e.g., suggest home remedies template to interpret body temperature), a new SPARQL query is done to return the description of one specific template. 
Each template enables building semantic-based WoT applications and indicates: 1) The sensors employed in the application, 2) the applicative domains, 3) the rules to semantically annotate data, 4) the rules to deduce new knowledge from data, 5) the domain knowledge comprised of ontologies and datasets to build domain-specific or cross-domain applications, and 6) the SPARQL query to request smarter data enriched with semantic web technologies and inferred data produced once the reasoning engine is executed.

Once the description of the template is returned by the SPARQL query engine, SWoT generator produces a template with interoperable ontologies, rules and datasets to: 1) Semantically annotate data, 2) run a reasoning mechanism over data, and 3) combine applicative domains by linking data. 
The main novelty of this research approach and tool is that developers do not need to be familiar with semantic web technologies. Moreover, another important aspect is that the SWoT generator will provide interoperability among semantic-based WoT applications.

\subsection{Machine-to-Machine Measurement Language}
\label{sec:m3language}
\textbf{M3 language} is used by the Semantic Annotator for unifying WoT data coming from heterogeneous projects, platforms and testbeds. The M3 language can be seen as a dictionary to describe and unify: 1) Sensor type, 2) unit, 3) sensor measurement type, and 4) applicative WoT domain \cite{gyrardglobecom2014}. The M3 language enables dealing with heterogeneous terms used in different projects such as synonyms. This language has been designed by extracting popular terms from LOV4IoT explained in section \ref{sec:lov4iot}. For instance, the M3 language deals with interoperability of terms when describing sensors: Precipitation sensor or rainfall sensor or delete ambiguities by explicitly adding the meaning of the data (e.g., body temperature differs from a room temperature). This is also an essential step to later easily interpret data and infer new information. Using the M3 compliant 'Semantic Annotator', data become compatible with the M3 language, an essential step for an easy interpretation of data. The M3 language has been implemented through the M3 ontology\footnote{\url{http://sensormeasurement.appspot.com/m3#}, Last visited: September 2016} (V1) or the M3-lite taxonomy (V2)\footnote{http://ontology.fiesta-iot.eu/ontologyDocs/fiesta-iot/doc, Last visited: September 2016}, more precisely they are extensions of the popular W3C Semantic Sensor Networks (SSN) ontology. The M3 ontology is mainly focused on the interoperability of data, by extending the \texttt{ssn:ObservationValue} concept and by providing subclasses of \texttt{ssn:Sensor} and \texttt{ssn:FeatureOfInterest}. M3 language is a cornerstone to enable interlinking cross-domain knowledge, a set of interoperable ontologies, datasets and rules reused by the SWoT generator to design cross-domain applications (e.g., smart home and weather forecasting).

\subsection{Sensor-Based Linked Open Rules}
\label{sec:slor}
\begin{figure}[!htbp]
\centering
\FIG{\includegraphics[width=\linewidth]{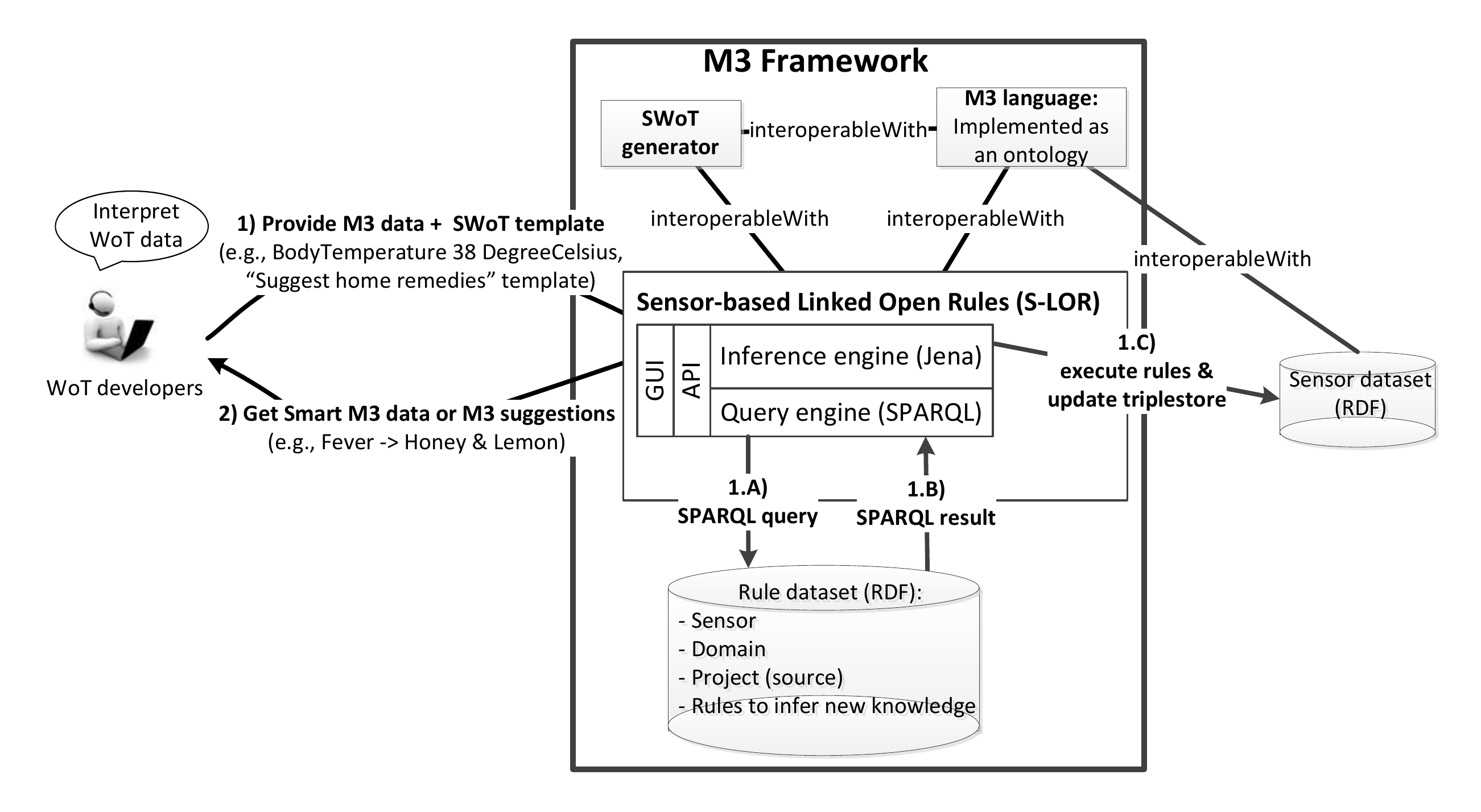}}
\caption{The S-LOR cycle for IoT data Interpretation}
\label{SLOR} 
\end{figure}

\textbf{Sensor Based Linked Open Rules (S-LOR)} is a sharing and reusing based approach, inspired from the Linked Open Data approach, to share the knowledge on the Web. S-LOR provides a dataset of interoperable rules used to infer new knowledge from WoT data (e.g., fever deduced from a body temperature) \cite{gyrardslor}. The rules have been written manually but extracted from the LOV4IoT dataset explained in section \ref{sec:lov4iot}. The rules are compliant with the M3 ontology mentioned previously and the Jena framework\footnote{https://jena.apache.org/}, more precisely the Jena inference engine\footnote{https://jena.apache.org/documentation/inference/}.  
Previously, the developer uses the M3 language to semantically annotate IoT data to produce M3-compliant data. Then, the data is enriched with new information thanks to the logic based reasoning engine S-LOR and the rules loaded that have been provided by the SWoT template as depicted in Figure \ref{SLOR}. Enriched M3 data is queried through a SPARQL query also provided by the SWoT template to provide domain-specific or cross-domains suggestions to the developer. For instance, S-LOR will deduce new information ``fever`` from the measurement body temperature 38$^\circ$C and will combine it to healthcare and naturopathy ontologies and datasets provided by the template. Finally, the developer will display results in a user-friendly interface, send alerts or even order actuators (e.g., open or close a door).

S-LOR has some limitations, it only deals with simple devices such as thermometer and not complicated sensors such as accelerometer or electrocardiograms (ECG).

\subsection{Linked Open Vocabularies for Internet of Things}
\label{sec:lov4iot}
To facilitate application development and its interoperability, the \textbf{Linked Open Vocabularies for Internet of Things (LOV4IoT)} dataset\footnote{http://sensormeasurement.appspot.com/?p=ontologies, Last visited: September 2016} has been designed \cite{gyrardlov4iotv1ficloud} \cite{gyrardlov4iotv2ficloud}. The main purpose of this dataset is to reuse the domain knowledge expertise already designed and available on the Web. LOV4IoT references 300 ontology-based projects exploiting sensors in various domains such as healthcare, building automation, food, agriculture, tourism, security, transportation, and smart city. Projects have been identified, studied and referenced since: 1) Devices are being used, 2) domain knowledge can be re-used in another context to design cross-domain use cases (e.g., the naturopathy application combines health, weather and smart kitchen domains), 3) ontologies and their evaluations have been designed, 4) ontologies or datasets can be re-used to ease interoperability and reduce development costs, 5) rule-based systems have been designed, 6) papers have been published in conferences or journals, and 7) the explanations of why semantic web technologies are integrated are provided.
However, the LOV4IoT dataset has some limitations since a lot of ontologies referenced lack of interoperability and best practices which hinder automation tasks. The LOV4IoT dataset has been mainly exploited to build the M3 language, SLOR interoperable rule dataset and the interoperable domain knowledge mainly exploited by the templates generated by the SWoT generator. A major challenge would be to automatically extract the domain knowledge from LOV4IoT and automatically re-design it to make it interoperable and enrich the SWoT generator with additional templates.

\subsection{Validation within FIESTA-IoT}
In the context of the FIESTA-IoT project (Federated Interoperable Semantic IoT/cloud Testbeds and Applications - introduced in Section \ref{sec:projects}) it is required to go beyond WoT. FIESTA-IoT focuses on interoperability of data, but also on the integration of testbeds (e.g., smart cities producing data) and experiments (e.g., WoT applications).
The data workflow that has been presented in Section \ref{sec:dataworkflow}, is being applied within FIESTA-IoT to address data interoperability issues. It is currently extended to design a generic approach, called SEG 3.0 methodology, to ensure data interoperability from data to end-user applications that could be applied on other domains than WoT \cite{gyrard2016methodology}. 
The SWoT generation is used to design and implement the concept ``Experiment-as-a Service`` (EaaS), mainly addressing interoperability of applications and services within IoT.
LOVIoT has been used to easily find  and reuse existing IoT ontologies to build the FIESTA-IoT ontology, which reuses, extends and aligns W3C SSN, IoT-Lite, and M3-lite ontologies. S-LOR is used to deduce meaningful information from data provided by testbeds (e.g., smart cities) registered within FIESTA-IoT.


\section{Summary and Future Work}
\label{sec:summary} 

\fakeparagraph{Summary} In this chapter, we presented a comprehensive study about the main WoT challenges, highlighting those that rely on combining heterogeneous IoT data for the design of smarter services and applications and that benefit from data interoperability. We discussed about the limitation of the existing work in the Web of Things and the justification about the need for a comprehensive framework. We presented the most relevant projects and tools working on semantic interoperability for annotating, linking and reasoning over Web of Things (WoT) data. To overcome some of the actual limitations of the existing WoT projects, the M3 framework has been designed and explained in detail. This chapter is mainly focused on assisting WoT developers in understanding the integration of semantic web technologies in order to deduce meaningful information from WoT data to build smarter WoT applications. 

\fakeparagraph{Future work}
Future activities focus on the refinement of the M3 framework and automatize as much as possible each component. For instance, LOV4IoT could be automatized to reuse, combine and extract the domain knowledge to easily deduce meaningful information from data. This knowledge is mainly reused to build interoperable semantic-based WoT applications and services provided by the SWoT generator. SWoT generator could be extended by investigating \textit{Semantic Web Services} and \textit{Linked Open Services} to enable the composition of simple services to provide more sophisticated WoT applications.

\begin{backmatter}

\begin{ack}[ACKNOWLEDGMENTS]

The activity presented in this chapter is partially funded by the European project "Federated Interoperable Semantic IoT/cloud Testbeds and Applications" (FIESTA-IoT) from the European Union's Horizon 2020 Programme with the Grant Agreement No. CNECT-ICT-643943. This chapter is extended from the authors’ past work particularly from the PhD thesis \cite{gyrard2015phdthesis} under the supervision of Prof. Christian Bonnet and Dr. Karima Boudaoud. We specially thank  Prof. Amit Sheth for providing feedback and valuable comments regarding this work.
\source{}
\end{ack}
\section*{REFERENCE}
\addcontentsline{toc}{section}{Reference}

\begin{thebibliography*}{10}

\providecommand{\url}[1]{{\tt #1}}
\providecommand{\urlprefix}{URL }
\expandafter\ifx\csname urlstyle\endcsname\relax
  \providecommand{\doi}[1]{doi:\discretionary{}{}{}#1}\else
  \providecommand{\doi}{doi:\discretionary{}{}{}\begingroup
  \urlstyle{rm}\Url}\fi
\providecommand{\bibinfo}[2]{#2}
\providecommand{\eprint}[2][]{\url{#2}}
\makeatletter\def\@biblabel#1{\bibinfo{label}{#1.}}\makeatother

\bibtype{book}%
\bibitem{aggarwal2013internet}
\bibinfo{author}{C. Aggarwal}, 
\bibinfo{author}{N. Ashish},
\bibinfo{author}{A. Sheth}, 
\bibinfo{title}{The internet of things: A survey from the data-centric perspective}.  
\bibinfo{year}{2013};
\bibinfo{edition}{Book: Managing and mining sensor data}.
\bibinfo{publisher}{Springer};
\bibinfo{pages}{383--428}.

\bibtype{book}%
\bibitem{guinard2010resource}
\bibinfo{author}{D. Guinard}, 
\bibinfo{author}{V. Trifa},
\bibinfo{author}{E. Wilde}, 
\bibinfo{title}{A resource oriented architecture for the web of things}.  
\bibinfo{year}{2010};
\bibinfo{edition}{Book: Internet of Things (IoT)}.
\bibinfo{publisher}{IEEE};
\bibinfo{pages}{1--8}.

\bibtype{book}%
\bibitem{vasseur2010interconnecting}
\bibinfo{author}{J.-P.. Vasseur}, 
\bibinfo{author}{A. Dunkels},
\bibinfo{title}{Interconnecting smart objects with ip: The next internet}.  
\bibinfo{year}{2010};
\bibinfo{publisher}{Morgan Kaufmann};

\bibtype{inproceedings}
\bibitem{guinard2011internet}
\bibinfo{title}{From the internet of things to the web of things: Resource-oriented architecture and best practices}.  
\bibinfo{author}{D. Guinard}, 
\bibinfo{author}{V. Trifa}, 
\bibinfo{author}{F. Mattern},
\bibinfo{author}{E. Wilde},
\bibinfo{booktitle}{From the internet of things to the web of things: Resource-oriented architecture and best practices}. 
\bibinfo{year}{2011};
\bibinfo{publisher}{Springer}.

\bibtype{inproceedings}
\bibitem{gangemi2009ontology}
\bibinfo{title}{Ontology design patterns}.  
\bibinfo{author}{A. Gangemi}, 
\bibinfo{author}{V. Presutti}, 
\bibinfo{booktitle}{Handbook on ontologies}. 
\bibinfo{year}{2009};
\bibinfo{publisher}{Springer}. 

\bibtype{inproceedings}
\bibitem{ipsn2017demo}
\bibinfo{title}{Demonstration Abstract: IoTSuite - A Framework to Design, Implement, and Deploy IoT Applications}.  
\bibinfo{author}{S. Chauhan}, 
\bibinfo{author}{P. Patel}, 
\bibinfo{author}{A. Sureka}, 
\bibinfo{author}{S. Chaudhary}, 
\bibinfo{booktitle}{15th ACM/IEEE International Conference on Information Processing in Sensor Networks (IPSN)}. 
\bibinfo{year}{2016};
\bibinfo{publisher}{IEEE}. 

\bibtype{inproceedings}
\bibitem{iotsuite-icse2016}
\bibinfo{title}{A Development Framework for Programming Cyber-physical Systems}.  
\bibinfo{author}{S. Chauhan}, 
\bibinfo{author}{P. Patel}, 
\bibinfo{author}{A. Sureka}, 
\bibinfo{author}{S. Chaudhary}, 
\bibinfo{booktitle}{Proceedings of the 2Nd International Workshop on Software Engineering for Smart Cyber-Physical Systems}. 
\bibinfo{year}{2016};
\bibinfo{publisher}{ACM}.

\bibtype{inproceedings}
\bibitem{iotsuite-icdcn}
\bibinfo{title}{IoTSuite: a ToolSuite for prototyping internet of things applications}.  
\bibinfo{author}{D. Soukaras}, 
\bibinfo{author}{P. Patel}, 
\bibinfo{author}{H. Song}, 
\bibinfo{author}{S. Chaudhary}, 
\bibinfo{booktitle}{The 4th International Workshop on Computing and Networking for Internet of Things (ComNet-IoT), co-located with 16th International Conference on Distributed Computing and Networking (ICDCN).}. 
\bibinfo{year}{2015};
\bibinfo{publisher}{Technical report}.

\bibtype{inproceedings}
\bibitem{leggieri2011incontext}
\bibinfo{title}{inContext-Sensing: LOD augmented sensor data?}.  
\bibinfo{author}{M. Leggieri}, 
\bibinfo{author}{A. Passant}, 
\bibinfo{author}{M. Hauswirth}, 
\bibinfo{booktitle}{Proceedings of the 10th International Semantic Web Conference (ISWC 2011)}. 
\bibinfo{year}{2011};
\bibinfo{publisher}{Springer}.

\bibtype{inproceedings}
\bibitem{gyrard2015wfiotBestPractices}
\bibinfo{title}{Semantic Web Methodologies, Best Practices and Ontology Engineering Applied to Internet of Things}.  
\bibinfo{author}{A. Gyrard}, 
\bibinfo{author}{M. Serrano}, 
\bibinfo{author}{G. Atemezing}, 
\bibinfo{booktitle}{WF-IOT 2015, World Forum on Internet of Things, 14-16 December 2015, Milan, Italy}. 
\bibinfo{year}{2015};
\bibinfo{publisher}{Springer}.

\bibtype{inproceedings}
\bibitem{scharffe2012enabling}
\bibinfo{title}{Enabling linkeddata publication with the datalift platform}.  
\bibinfo{author}{F. Scharffe}, 
\bibinfo{author}{G. Atemezing}, 
\bibinfo{author}{R. Troncy}, 
\bibinfo{author}{F. Gandon}, 
\bibinfo{author}{S. Villata}, 
\bibinfo{author}{and others}, 
\bibinfo{booktitle}{Proc. AAAI workshop on semantic cities}. 
\bibinfo{year}{2012};

\bibtype{misc}%
\bibitem{suarez2010neon}
\bibinfo{author}{M.-C. Suarez-Figueroa}, 
\bibinfo{title}{NeOn Methodology for building ontology networks: specification, scheduling and reuse}.  
\bibinfo{year}{2010};

\bibtype{misc}%
\bibitem{noy2001ontology}
\bibinfo{author}{N. Noy}, 
\bibinfo{author}{D. McGuinness}, 
\bibinfo{title}{Ontology development 101: A guide to creating your first ontology}.
\bibinfo{publisher}{Stanford knowledge systems laboratory technical report KSL-01-05 and Stanford medical informatics technical report SMI-2001-0880};
\bibinfo{year}{2001};

\bibtype{misc}
\bibitem{gyrard2015phdthesis}
\bibinfo{author}{A. Gyrard}, 
\bibinfo{title}{Designing Cross-Domain Semantic Web of Things Applications}.  
\bibinfo{year}{2015};
\bibinfo{publisher}{Eurecom};

\bibtype{misc}
\bibitem{henson2013semantics}
\bibinfo{author}{C. Henson}, 
\bibinfo{title}{A Semantics-based Approach to Machine Perception}.  
\bibinfo{year}{2013};
\bibinfo{publisher}{Wright State University}; 

\bibtype{misc}
\bibitem{patel-evalutioniot}
\bibinfo{author}{P. Patel, A. Kattepur, D. Cassou, and  G. Bouloukakis}, 
\bibinfo{title}{{Evaluating the Ease of Application Development for the Internet of Things}}.  
\bibinfo{year}{2013};
\bibinfo{publisher}{INRIA-Paris};  

\bibtype{misc}
\bibitem{iswc2016tutorial}
\bibinfo{author}{A. Gyrard, P.Patel, S. Dutta, M. Ali}, 
\bibinfo{title}{Semantic web meets internet of things (iot) and web of things (wot).},  
\bibinfo{booktitle}{The 15th International Conference on Semantic Web }. 
\bibinfo{year}{2016};
\bibinfo{publisher}{CSUR-WS}; 

\bibtype{Article}
\bibitem{swotsuiteframework}
\bibinfo{title}{SWoTSuite: {A} Development Framework for Prototyping Cross-domain
               Semantic Web of Things Applications}.  
\bibinfo{author}{P. Patel}, 
\bibinfo{author}{A. Gyrard},
\bibinfo{author}{D. Thakkar}, 
\bibinfo{author}{A. Sheth}, 
\bibinfo{author}{M. Serrano}, 
\bibinfo{journal}{The 15th International Conference on Semantic Web (ISWC)} 
\bibinfo{year}{2016};
\bibinfo{publisher}{arxiv};
\bibinfo{pages}{8}. 

\bibtype{Article}
\bibitem{gpce2015}
\bibinfo{title}{Evaluating a Development Framework for Engineering
Internet of Things Applications}.  
\bibinfo{author}{P. Patel}, 
\bibinfo{author}{T. Luo},
\bibinfo{author}{U. bellur}, 
\bibinfo{year}{2015};
\bibinfo{publisher}{arxiv};
\bibinfo{pages}{13}.

\bibtype{Article}
\bibitem{barnaghi2012semantics}
\bibinfo{title}{Semantics for the Internet of Things: early progress and back to the future}.  
\bibinfo{author}{P. Barnaghi}, 
\bibinfo{author}{W. Wang},
\bibinfo{author}{C. Henson}, 
\bibinfo{author}{K. Taylor}, 
\bibinfo{journal}{International Journal on Semantic Web and Information Systems (IJSWIS)} 
\bibinfo{year}{2012};
\bibinfo{volume}{8}(\bibinfo{number}{1}):
\bibinfo{publisher}{IGI Global};
\bibinfo{pages}{1--21}.

\bibtype{Article}%
\bibitem{jara2014semantic}
\bibinfo{title}{Semantic Web of Things: an analysis of the application semantics for the IoT moving towards the IoT convergence}.  
\bibinfo{author}{A. Jara}, 
\bibinfo{author}{A. Olivieri},
\bibinfo{author}{Y. Bocchi}, 
\bibinfo{author}{M. Jung}, 
\bibinfo{author}{W. Kastner}, 
\bibinfo{author}{A. Skarmeta}, 
\bibinfo{journal}{International Journal of Web and Grid Services} 
\bibinfo{year}{2014};
\bibinfo{volume}{10}(\bibinfo{number}{2}):
\bibinfo{publisher}{Inderscience};
\bibinfo{pages}{244--272}.

\bibtype{Article}%
\bibitem{compton2012ssn}
\bibinfo{title}{The ssn ontology of the w3c semantic sensor network incubator group}.  
\bibinfo{author}{M. Compton}, 
\bibinfo{author}{P. Barnaghi},
\bibinfo{author}{L. Bermudez}, 
\bibinfo{author}{R. Garcia-Castro}, 
\bibinfo{author}{C. Henson}, 
\bibinfo{author}{A. Herzog}, 
\bibinfo{author}{and others}, 
\bibinfo{journal}{Web Semantics: Science, Services and Agents on the World Wide Web} 
\bibinfo{year}{2012};
\bibinfo{publisher}{Elsevier}.

\bibtype{Article}%
\bibitem{botts2008ogc}
\bibinfo{title}{OGC sensor web enablement: Overview and high level architecture}.  
\bibinfo{author}{M. Botts}, 
\bibinfo{author}{G. Percivall},
\bibinfo{author}{C. Reed}, 
\bibinfo{author}{J. Davidson}, 
\bibinfo{journal}{GeoSensor networks} 
\bibinfo{year}{2008};
\bibinfo{pages}{175--190}.
\bibinfo{publisher}{Springer}.

\bibtype{Article}
\bibitem{vandenbusschelov2015}
\bibinfo{title}{LOV: a gateway to reusable semantic vocabularies on the Web}.  
\bibinfo{author}{P.-Y. Vandenbussche}, 
\bibinfo{author}{G. Atemezing},
\bibinfo{author}{M. Poveda-Villalon}, 
\bibinfo{author}{B. Vatant}, 
\bibinfo{journal}{Semantic Web Journal} 
\bibinfo{year}{2015};

\bibtype{Article}%
\bibitem{zeng2011web}
\bibinfo{title}{The web of things: A survey}.  
\bibinfo{author}{D. Zeng}, 
\bibinfo{author}{S. Guo},
\bibinfo{author}{Z. Cheng}, 
\bibinfo{journal}{Journal of Communications} 
\bibinfo{year}{2011};
\bibinfo{volume}{6}(\bibinfo{number}{6}):
\bibinfo{pages}{424--438}.

\bibtype{inproceedings}
\bibitem{gyrard2016methodology}
\bibinfo{author}{A. Gyrard}, 
\bibinfo{author}{{M. Serrano}},
\bibinfo{title}{Connected Smart Cities: Interoperability with SEG 3.0 for the Internet of Things}, 
\bibinfo{booktitle}{30th IEEE International Conference on Advanced Information Networking and Applications Workshops, 2016, Crans-Montana, Switzerland.}. 
\bibinfo{bknote}{23-25 March 2016, Crans-Montana, Switzerland.}. 
\bibinfo{year}{2016};

\bibtype{inproceedings}
\bibitem{gyrard2014wfiot}
\bibinfo{author}{A. Gyrard}, 
\bibinfo{author}{{C. Bonnet}},
\bibinfo{author}{K. Boudaoud}, 
\bibinfo{title}{Enrich machine-to-machine data with semantic web technologies for cross-domain applications}, 
\bibinfo{booktitle}{WF-IOT 2014, World Forum on Internet of Things, }. 
\bibinfo{bknote}{6-8 March 2014, Seoul, Korea}. 
\bibinfo{year}{2014};

\bibtype{Article}
\bibitem{petrolo2015towards}
\bibinfo{title}{Towards a smart city based on cloud of things, a survey on the smart city vision and paradigms}.  
\bibinfo{author}{R. Petrolo}, 
\bibinfo{author}{V. Loscri},
\bibinfo{author}{N. Mitton}, 
\bibinfo{journal}{Transactions on Emerging Telecommunications Technologies} 
\bibinfo{year}{2015};
\bibinfo{publisher}{Wiley Online Library}.

\bibtype{Article}
\bibitem{sheth2008semantic}
\bibinfo{author}{A. Sheth}, 
\bibinfo{author}{{C. Henson}},
\bibinfo{author}{S.S. Sahoo}, 
\bibinfo{title}{Semantic sensor web}, 
\bibinfo{journal}{Internet Computing}. 
\bibinfo{publisher}{IEEE}.
\bibinfo{pages}{78--83}.
\bibinfo{volume}{12}(\bibinfo{number}{4}):
\bibinfo{year}{2008};

\bibtype{inproceedings}
\bibitem{henson2009semsos}
\bibinfo{title}{henson2009semsos}.  
\bibinfo{author}{C. Henson}, 
\bibinfo{author}{J. Pschorr}, 
\bibinfo{author}{A. Sheth},
\bibinfo{author}{K. Thirunarayan},
\bibinfo{booktitle}{Collaborative Technologies and Systems, 2009. CTS'09. International Symposium on}. 
\bibinfo{year}{2009};
\bibinfo{pages}{44--53}.
\bibinfo{publisher}{IEEE}.

\bibtype{inproceedings}
\bibitem{patni2010linked}
\bibinfo{title}{Linked sensor data}.  
\bibinfo{author}{H. Patni},
\bibinfo{author}{C. Henson}, 
\bibinfo{author}{A. Sheth},
\bibinfo{booktitle}{Collaborative Technologies and Systems (CTS), 2010 International Symposium on}. 
\bibinfo{year}{2010};
\bibinfo{pages}{362--370}.
\bibinfo{publisher}{IEEE}. 

\bibtype{inproceedings}
\bibitem{apsec2015}
\bibinfo{title}{Building Enterprise-Grade Internet of Things Applications}.  
\bibinfo{author}{P. Patel},
\bibinfo{author}{V. Kaulgud}, 
\bibinfo{author}{P. Chandra},
\bibinfo{author}{A. Kumar},
\bibinfo{booktitle}{2015 Asia-Pacific Software Engineering Conference (APSEC)}. 
\bibinfo{year}{2015};
\bibinfo{pages}{4-5}.
\bibinfo{publisher}{IEEE}. 

\bibtype{Article}
\bibitem{bizer2009linked}
\bibinfo{title}{Linked data-the story so far}.  
\bibinfo{author}{C. Bizer}, 
\bibinfo{author}{T. Heath},
\bibinfo{author}{T. Berners-Lee},  
\bibinfo{journal}{International Journal on Semantic Web and Information Systems (IJSWIS)} 
\bibinfo{year}{2009};
\bibinfo{volume}{5}(\bibinfo{number}{3}):
\bibinfo{publisher}{IGI Global};
\bibinfo{pages}{1--22}.

\bibtype{Article}
\bibitem{pfisterer2011spitfire}
\bibinfo{title}{SPITFIRE: toward a semantic web of things}.  
\bibinfo{author}{D. Pfisterer}, 
\bibinfo{author}{K. Romer},
\bibinfo{author}{D. Bimschas}, 
\bibinfo{author}{O. Kleine}, 
\bibinfo{author}{R. Mietz}, 
\bibinfo{author}{C. Truong}, 
\bibinfo{author}{H. Hasemann}, 
\bibinfo{author}{A. Kroller}, 
\bibinfo{author}{M. Pagel}, 
\bibinfo{author}{M. Hauswirth}, 
\bibinfo{journal}{Communications Magazine, IEEE} 
\bibinfo{year}{2011};
\bibinfo{volume}{49}(\bibinfo{number}{11}):
\bibinfo{publisher}{IEEE};
\bibinfo{pages}{40--48}.

\bibtype{soldatos2015openiot}
\bibitem{soldatos2015openiot}
\bibinfo{title}{OpenIoT: Open Source Internet-of-Things in the Cloud}.
\bibinfo{author}{J. Soldatos},
\bibinfo{author}{N. Kefalakis},
\bibinfo{author}{M. Hauswirth},
\bibinfo{author}{M. Serrano},
\bibinfo{author}{J.-P. Calbimonte}, 
\bibinfo{author}{M. Riahi}, 
\bibinfo{author}{K. Aberer},
\bibinfo{author}{P. Jayaraman},
\bibinfo{author}{A. Zaslavsky},
\bibinfo{author}{I. Zarko},
\bibinfo{booktitle}{Interoperability and Open-Source Solutions for the Internet of Things}. 
\bibinfo{year}{2015};
\bibinfo{publisher}{Springer}.
\bibinfo{pages}{13--25}.

\bibtype{inproceedings}
\bibitem{barnaghicitypulse}
\bibinfo{title}{CityPulse: Real-Time IoT Stream Processing and Large-scale Data Analytics for Smart City Applications}.  
\bibinfo{author}{P. Barnaghi}, 
\bibinfo{author}{R. Tonjes}, 
\bibinfo{author}{J. Holler},
\bibinfo{author}{M. Hauswirth},
\bibinfo{author}{A. Sheth},
\bibinfo{author}{P. Anantharam},
\bibinfo{booktitle}{Europen Semantic Web Conference (ESWC) 2014}. 
\bibinfo{year}{2014};
\bibinfo{publisher}{Springer}.

\bibtype{Article}
\bibitem{petrolo2015towards}
\bibinfo{title}{Towards a smart city based on cloud of things, a survey on the smart city vision and paradigms}.  
\bibinfo{author}{R. Petrolo}, 
\bibinfo{author}{V. Loscri},
\bibinfo{author}{N. Mitton}, 
\bibinfo{journal}{Transactions on Emerging Telecommunications Technologies} 
\bibinfo{year}{2015};
\bibinfo{publisher}{Wiley Online Library}.

\bibtype{inproceedings}
\bibitem{petrolo2014integrating}
\bibinfo{title}{Integrating wireless sensor networks within a city cloud}.  
\bibinfo{author}{R. Petrolo}, 
\bibinfo{author}{N. Mitton}, 
\bibinfo{author}{J. Soldatos},
\bibinfo{author}{M. Hauswirth},
\bibinfo{author}{G. Schiele},
\bibinfo{booktitle}{Sensing, Communication, and Networking Workshops (SECON Workshops, 2014 Eleventh Annual IEEE International Conference on}. 
\bibinfo{year}{2014};
\bibinfo{publisher}{IEEE}.
\bibinfo{pages}{24--27}.

\bibtype{inproceedings}
\bibitem{gyrard2014stac}
\bibinfo{title}{An Ontology-Based Approach for Helping to Secure the ETSI Machine-to-Machine Architecture}.  
\bibinfo{author}{A. Gyrard}, 
\bibinfo{author}{C. Bonnet}, 
\bibinfo{author}{K. Boudaoud},
\bibinfo{booktitle}{IEEE International Conference on Internet of Things 2014 (iThings)}. 
\bibinfo{year}{2014};
\bibinfo{publisher}{IEEE}.

\bibtype{inproceedings}
\bibitem{gyrard2015m3}
\bibinfo{title}{Cross-domain Internet of Things application development: M3 framework and evaluation}.  
\bibinfo{author}{A. Gyrard}, 
\bibinfo{author}{S. Datta}, 
\bibinfo{author}{C. Bonnet}, 
\bibinfo{author}{K. Boudaoud},
\bibinfo{booktitle}{FICLOUD 2015, 3rd International Conference on Future Internet of Things and Cloud}. 
\bibinfo{year}{2015};
\bibinfo{bknote}{August 24-26, 2015, Rome, Italy}. 
\bibinfo{publisher}{IEEE}.

\bibtype{inproceedings}
\bibitem{gyrard2015swotgenerator}
\bibinfo{title}{Assisting IoT Projects and Developers in Designing Interoperable Semantic Web of Things Applications}.  
\bibinfo{author}{A. Gyrard}, 
\bibinfo{author}{C. Bonnet}, 
\bibinfo{author}{K. Boudaoud},
\bibinfo{author}{M. Serrano},
\bibinfo{booktitle}{IEEE International Conference on Internet of Things 2015 (iThings)}. 
\bibinfo{year}{2015};
\bibinfo{publisher}{IEEE}.

\bibtype{inproceedings}
\bibitem{gyrardslor}
\bibinfo{title}{Helping IoT application developers with sensor-based linked open rules}.  
\bibinfo{author}{A. Gyrard}, 
\bibinfo{author}{C. Bonnet}, 
\bibinfo{author}{K. Boudaoud},
\bibinfo{booktitle}{SSN 2014, 7th International Workshop on Semantic Sensor Networks in conjunction with the 13th International Semantic Web Conference (ISWC 2014), 19-23 October 2014, {R}iva Del Garda, Italy}. 
\bibinfo{year}{2014};
\bibinfo{publisher}{Springer}.

\bibtype{Article}
\bibitem{ganzKat2014}
\bibinfo{title}{Automated Semantic Knowledge Acquisition From Sensor Data}.  
\bibinfo{author}{F. Ganz}, 
\bibinfo{author}{P. Barnaghi},
\bibinfo{author}{F. Carrez}, 
\bibinfo{Journal}{IEEE Systems} 
\bibinfo{year}{2016};
\bibinfo{publisher}{IEEE}.

\bibtype{Article}
\bibitem{patel-jss}
\bibinfo{title}{Enabling high-level application development for the Internet of Things}.  
\bibinfo{author}{P. Patel}, 
\bibinfo{author}{D. Cassou},
\bibinfo{Journal}{Journal of Systems and Software} 
\bibinfo{year}{2015};
\bibinfo{publisher}{ScienceDirect, Elsevier}.

\bibtype{inproceedings}
\bibitem{prazeres2016FogOfThings}
\bibinfo{title}{SOFT-IoT:Self-OrganizingFOGofThings}.  
\bibinfo{author}{C. Prazeres}, 
\bibinfo{author}{M. Serrano}, 
\bibinfo{booktitle}{30th IEEE International Conference on Advanced Information Networking and Applications (IEEE AINA-2016) workshop on Pervasive Internet of Things and Smart Cities (PITSaC)}. 
\bibinfo{year}{2016};
\bibinfo{publisher}{ACM}.

\bibtype{inproceedings}
\bibitem{tummarello2007sindice}
\bibinfo{title}{Sindice.com: Weaving the open linked data}.  
\bibinfo{author}{G. Tummarello}, 
\bibinfo{author}{R. Delbru}, 
\bibinfo{author}{E. Oren}, 
\bibinfo{year}{2007};
\bibinfo{publisher}{Springer}.

\bibtype{inproceedings}
\bibitem{tummarello2007sindice}
\bibinfo{title}{Sindice.com: Weaving the open linked data}.  
\bibinfo{author}{G. Tummarello}, 
\bibinfo{author}{R. Delbru}, 
\bibinfo{author}{E. Oren}, 
\bibinfo{year}{2007};
\bibinfo{publisher}{Springer}.

\bibtype{inproceedings}
\bibitem{calbimonte2014xgsn}
\bibinfo{title}{XGSN: An Open-source Semantic Sensing Middleware for the Web of Things}.  
\bibinfo{author}{J-P. Calbimonte}, 
\bibinfo{author}{S. Sarni}, 
\bibinfo{author}{J. Eberle}, 
\bibinfo{author}{K. Aberer}, 
\bibinfo{year}{2014};
\bibinfo{booktitle}{7th International Workshop on Semantic Sensor Network}. 
\bibinfo{publisher}{Springer}.

\bibtype{inproceedings}
\bibitem{lePhuoc2014GraphOfThings}
\bibinfo{title}{Enabling Live Exploration on The Graph of Things?}.  
\bibinfo{author}{D. Le-Phuoc}, 
\bibinfo{author}{H. Quoc}, 
\bibinfo{author}{Q. Ngo}, 
\bibinfo{author}{T. Nhat}, 
\bibinfo{author}{M. Hauswirth}, 
\bibinfo{year}{2014};
\bibinfo{booktitle}{International Semantic Web Conference - Proceedings of the Semantic Web Challenge}. 
\bibinfo{publisher}{Springer}.

\bibtype{inproceedings}
\bibitem{ding2004swoogle}
\bibinfo{title}{Swoogle: a search and metadata engine for the semantic web}.  
\bibinfo{author}{L. Ding}, 
\bibinfo{author}{T. Finin}, 
\bibinfo{author}{T. Finin}, 
\bibinfo{author}{A. Joshi}, 
\bibinfo{author}{R. Pan}, 
\bibinfo{author}{R. Cost}, 
\bibinfo{author}{Y. Peng}, 
\bibinfo{author}{P. Reddivari}, 
\bibinfo{author}{V. Doshi}, 
\bibinfo{author}{J. Sachs}, 
\bibinfo{booktitle}{Proceedings of the thirteenth ACM international conference on Information and knowledge management}. 
\bibinfo{year}{2004};
\bibinfo{pages}{652--659}.
\bibinfo{publisher}{ACM}.

\bibtype{inproceedings}
\bibitem{gyrardlov4iotv1ficloud}
\bibinfo{title}{LOV4IoT: A second life for ontology-based domain knowledge to build Semantic Web of Things applications}.  
\bibinfo{author}{A. Gyrard}, 
\bibinfo{author}{C. Bonnet}, 
\bibinfo{author}{K. Boudaoud}, 
\bibinfo{author}{M. Serrano}, 
\bibinfo{year}{2016};
\bibinfo{booktitle}{4rd International Conference on Future Internet of Things and Cloud (FiCloud 2016), 22-24 August 2016, Vienna, Austria}. 
\bibinfo{publisher}{IEEE}.

\bibtype{inproceedings}
\bibitem{gyrardlov4iotv2ficloud}
\bibinfo{title}{Reusing and Unifying Background Knowledge for Internet of Things with LOV4IoT}.  
\bibinfo{author}{A. Gyrard}, 
\bibinfo{author}{G. Atemezing}, 
\bibinfo{author}{C. Bonnet}, 
\bibinfo{author}{K. Boudaoud}, 
\bibinfo{author}{M. Serrano}, 
\bibinfo{year}{2016};
\bibinfo{booktitle}{4rd International Conference on Future Internet of Things and Cloud (FiCloud 2016), 22-24 August 2016, Vienna, Austria}. 
\bibinfo{publisher}{IEEE}.

\bibtype{inproceedings}
\bibitem{gyrardglobecom2014}
\bibinfo{title}{Standardizing Generic Cross-Domain Applications in Internet of Things}.  
\bibinfo{author}{A. Gyrard}, 
\bibinfo{author}{S.K. Datta}, 
\bibinfo{author}{C. Bonnet}, 
\bibinfo{author}{K. Boudaoud}, 
\bibinfo{year}{2014};
\bibinfo{booktitle}{Third Workshop on Telecommunications Standards, Part of IEEE Globecom 2014, Austin, TX, USA, 8-12 December 2014}. 
\bibinfo{publisher}{IEEE}.


\end{thebibliography*}

\end{backmatter}

\end{document}